\documentclass[11pt,a4paper]{article}

\usepackage[T1]{fontenc}
\usepackage[utf8]{inputenc}

\usepackage{graphicx}

\usepackage[margin=1.5in]{geometry}

\renewcommand{\normalsize}{\fontsize{11pt}{12pt}\selectfont}
\renewcommand{\Huge}{\fontsize{18pt}{22pt}\selectfont}

\usepackage{fancyhdr}

\fancypagestyle{plain}{
    \fancyhf{}
    \fancyhead[R]{}
    \fancyfoot[R]{{ \textit{SIGBPS 2018 Workshop on Blockchain and Smart Contract, San Francisco}}\hspace{5mm}\textbf{\thepage}}
  }

\usepackage[hidelinks]{hyperref}

\usepackage{tabularx}  

\usepackage{amsmath}
\usepackage{amssymb}
\usepackage{amsthm}

\usepackage{url}

\usepackage{enumitem}
\setitemize{noitemsep,topsep=2pt,parsep=4pt,partopsep=2pt}
\setenumerate{noitemsep,topsep=2pt,parsep=4pt,partopsep=2pt}
\setdescription{noitemsep,topsep=2pt,parsep=4pt,partopsep=2pt}

\newcommand{\DM} {\normalfont{\textsf{DM}}}
\newcommand{\dDM} {\normalfont{\textsf{dDM}}}
\newcommand{\dPDM} {\normalfont{\textsf{dPDM}}}

\newcommand{\calA}{\ensuremath{\mathcal{A}}}
\newcommand{\calB}{\ensuremath{\mathcal{B}}}
\newcommand{\calL}{\ensuremath{\mathcal{L}}}
\newcommand{\calN}{\ensuremath{\mathcal{N}}}
\newcommand{\calR}{\ensuremath{\mathcal{R}}}
\newcommand{\calS}{\ensuremath{\mathcal{S}}}

\theoremstyle{definition}
\newtheorem{definition}{Definition}[section]

\title{\Huge\bfseries Wibson: A Decentralized Data Marketplace}

\author{  \begin{tabularx}{\textwidth}{@{}*2{>{\centering\arraybackslash}X}@{}}
    \textbf{Matias Travizano}        
    & \textbf{Carlos Sarraute} \\
    mat@wibson.org & charles@wibson.org \\
    \\
    \textbf{Gustavo Ajzenman}        
    & \textbf{Martin Minnoni} \\
    gustavo@wibson.org & martin@wibson.org \\
  \end{tabularx}
}

\date{}

\begin{document}

\maketitle


\begin{abstract}
{\normalsize  
\noindent
{Our aim is for Wibson to be a blockchain-based, decentralized data marketplace that provides individuals a way to securely and anonymously sell information in a trusted environment. The combination of the Wibson token and blockchain-enabled smart contracts hopes to allow Data Sellers and Data Buyers to transact with each other directly while providing individuals the ability to maintain anonymity as desired.}

\medskip
\noindent
{Wibson intends that its data marketplace will provide infrastructure and financial incentives for individuals to securely sell personal information without sacrificing personal privacy. Data Buyers receive information from willing and actively participating individuals with the benefit of knowing that the personal information should be accurate and current.}

}
\end{abstract}



\section{Introduction}
\label{sec:introduction}

We propose a new market-based approach that leverages the latest developments in blockchain, cryptography and market design to level the playing field between data consumers (companies, organizations) and data owners (essentially, individuals).

{Wibson is a blockchain-based, decentralized data marketplace} that provides the infrastructure for individuals to securely and anonymously sell personal information that is validated for accuracy~\cite{wibson2018}. Wibson is built up on a set of core principles: {transparency, anonymity, fairness, censorship resistant, and the individual's ultimate control over the use of their personal information.}

The design and price of information in data markets is an active field of study~\cite{bergemann2018markets,bergemann2018design}. In the Wibson marketplace that we describe here, citizens will be able to participate in an efficiently functioning, decentralized marketplace that provides both financial incentives and control over their personal information, all without sacrificing privacy.


\section{Marketplace Definitions}
\label{sec:definition-dpdm}

\begin{definition}
A \emph{Data Marketplace} (\DM) is a platform for the trade of information which provides:
\begin{itemize}
\item The infrastructure for a free market whereby parties engage in exchange: sellers offering their data in exchange for money from buyers.
\item Allow any data tradeable item to be evaluated and valued.
\item Incentives for all players to be honest, and an enforcement system to take actions if a dishonest behavior is chosen.
\item Incentives for all players to ensure that data is trustworthy, and to provide quality data with the addition of an enforcement system to ensure this.
\end{itemize}
\end{definition}

\begin{definition}
A \emph{Decentralized Data Marketplace} (\dDM) is a {\DM} such that:
\begin{itemize}
\item There is no central authority which regulates the participants of the market.
\item There is no central data repository. The users who generate the data are the owners of the data, and keep their data in their own devices having full control of their data assets. 
\item There is no central funds repository, therefore providing a trustless system where actors do not have to entrust their funds to a third party.
\end{itemize}
\end{definition}

\begin{definition}
A \emph{Decentralized Privacy-Preserving Data Marketplace} (\dPDM) is a {\dDM} which allows users to sell private information, while providing them the following privacy guarantees:
\begin{itemize}
\item Participants anonymity: the identity of the Sellers and Buyers is not revealed, without their consent. In particular, the identity of the Data Seller is not revealed to the Data Buyer, without the consent of the Data Seller.
\item Transparency over Data usage: the Data Seller has always visibility on how his Data is used by the Buyer.
\item Control over Data usage: the Data Seller can modify the rights over its Data at any time.

\end{itemize}
\end{definition}


\section{Wibson: a Decentralized Data Marketplace}
\label{sec:wibson-protocol}

In this section, we present the Wibson Decentralized Data Marketplace,
which is inspired by the {\dPDM} definitions, while allowing for trade-offs that are practical and necessary to ensure a market adapted to today's industry structure. 

\subsection{Marketplace participants}
\label{sec:participants}

The protocol is conducted by three types of market participants: Data Sellers, Data Buyers, and Notaries. 
\begin{description}
\item[Data Seller:] The Seller owns data and has rights to sell that data. The typical case is an individual selling his/her personal data. 
\item[Data Buyer:] Any entity who wants to purchase data. Typically the Buyer is an organization intending to train Machine Learning algorithms and models by using the data acquired.
\item[Notary:] We introduce the role of Notary as a verification system to verify participants information when required, verify data quality and trustworthiness when required, and arbitrate in case of conflict between 
Data Sellers and Data Buyers. 

To qualify as a Notary, the Notary must have access to \emph{ground truth} information with respect to the data being exchanged in the marketplace. 
In other words, the Notary will be an entity with information from their own files on the Data Sellers and will be able to verify it.

\end{description}

\noindent
We say that the marketplace is \textbf{decentralized} since any participant which qualifies
can enter the marketplace as Data Seller, Data Buyer or Notary.
There is no central authority which controls the participation in the market, or gives/denies permission to act in the market. This is a clear contrast with most common marketplaces, controlled by a central authority.

\subsection{Protocol description}
\label{sec:protocol-description}

\begin{figure}[ht]
	\centering
	{\includegraphics[width=\linewidth]{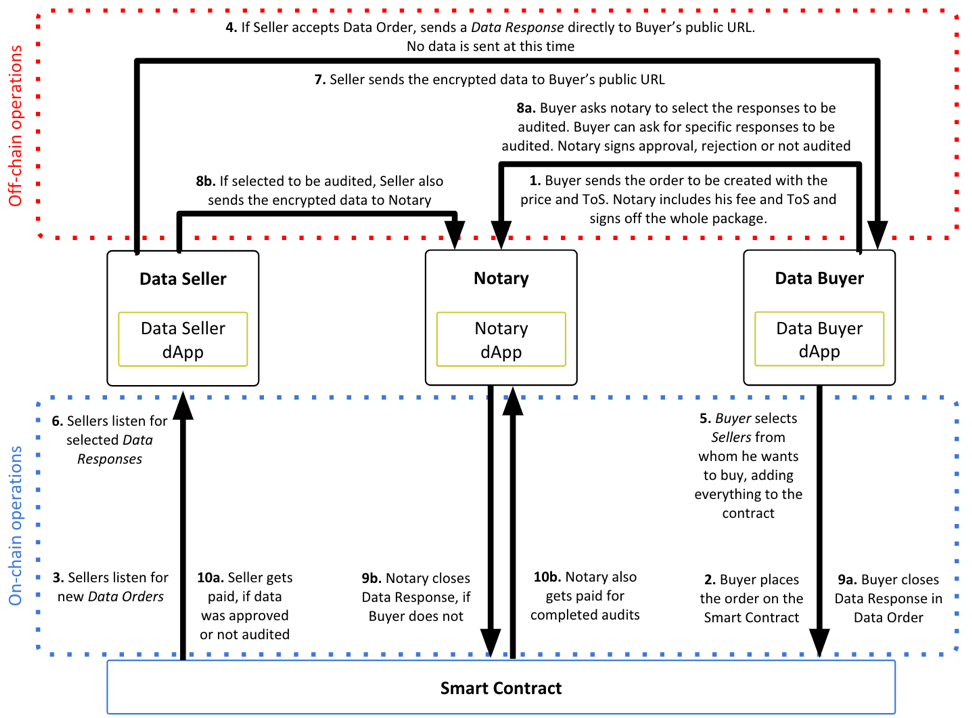}}
	\caption{Wibson's normal flow of operations.} 
	\label{fig:normal-flow}
\end{figure}

We present here the mechanisms and flow of operations of the Wibson protocol. 
The normal flow of operations is represented in Figure~\ref{fig:normal-flow}.

\begin{enumerate}

\item The Data Buyer $\calB$ creates a Data Order query 
$ \mbox{\textsf{DO}} = \left< \calA, \calR, 
PK_{\calB}, U_{\calB}, m_a, \mbox{\textsf{tc}} \right>$ 
and sends it to all the Notaries he wants to include, in order to obtain their fees, terms of service for the $ \mbox{\textsf{DO}}$ and their signatures over all this information in agreement with the conditions described by the Data Buyer.
The $ \mbox{\textsf{DO}}$ includes:
\begin{enumerate}[label=(\roman*)]
\item intended audience $\calA$ (filter of potential sellers), 
\item data requested $\calR$, 
\item the Data Buyer's public key $PK_{\calB}$,
\item public URL to upload Data Seller's responses and encrypted data via HTTPS post $U_{\calB}$,
\item minimum audit budget $m_a$, 
\item terms and conditions of data use $\mbox{\textsf{tc}}$.
\end{enumerate}

\item The Data Buyer sends the Data Order to the smart contract
adding also the list of notaries with their fees, terms of service and signatures
$\calL = \{ \calN_{k_1}, \ldots, \calN_{k_s} \}$. The Data Buyer also sets the price $p$ that he will pay for each Data Response to the Data Order.
The tokens (corresponding to minimum audit budget $m_a$) leave the Data Buyer's control.

\item Sellers monitor Data Orders and look for opportunities where they:
\begin{enumerate}[label=(\roman*)]
\item match Data Order's audience $\calA$, 
\item agree on data requested $\calR$, 
\item accept suggested Notary (only one),
\item accept the Data Order price $p$,
\item accept terms and conditions of data use.
 \end{enumerate}

\item Data Sellers may send Data Responses to Data Buyer's public URL (off-chain).

Data Response 
$\mbox{\textsf{DR}} = \left< E_i, \textsf{DO}, p, \sigma_i, H(\sigma_i || D_i),   
\calN^*_i, \mbox{\textsf{tc}}, \mbox{\textsf{sig}}_{\calS_i} \right> $ includes:
\begin{enumerate}[label=(\roman*)]
\item address $E_i$ to receive the payments, 
\item the Data Order's address $\textsf{DO}$,
\item the price $p$ of the Data Order (for which they are willing to sell requested data), 
\item hash of unencrypted data $D_i$ prepended with a random salt $\sigma_i$, 
computed using SHA-256~\cite{lilly2004device}:
$\left< \sigma_i, H(\sigma_i || D_i) \right> , $
\item selected Notary $\calN^*_i \in \calL$ who is included in the list of Notaries $\calL$ who have signed. 
\item link to the accepted terms and conditions of data use $\mbox{\textsf{tc}}$,
\item the Data Seller's signature over all of the above  $\mbox{\textsf{sig}}_{\calS_i}$ (to add in the blockchain).
\end{enumerate}

No Data Seller data is sent at this point.

\item Data Buyer $\calB$ chooses the set of Data Sellers $T$ from whom he wants to buy and adds them to the Data Order contract.  
Tokens to pay Data Sellers leave the Data Buyer's control at this point (the selection happens on-chain in the transaction).
The Data Buyer also pays for eventual audits, if the minimum budget for audit $m_a$ was already depleted.


\item Data Sellers listen for approved Data Responses.

\item Data Sellers $\calS_i \in T$ who have their offer approved upload the data file (encrypted with the public key of the Data Buyer $PK_{\calB}$) to the requested address $U_{\calB}$.

\item Once the Data Buyer receives the personal information, the next step is to close the transaction and transfer the tokens accordingly. 

In order to close a Data Response over a Data Order, the Data Buyer must check with the Notary if the data must be notarized or not. Then the Notary will hand over to the Data Buyer a signed certificate specifying one of the following scenarios:
(a) the data will not be notarized;
(b) the data was notarized and is valid;
(c) the data was notarized and is invalid.

Unless the Data Buyer specifically requests the Notary to audit a Data Seller, it is up to the Notary to decide which Data Sellers are going to be notarized.


\item With the Notary's signed certificate, the Data Buyer can close the given Data Response (a.k.a. Data Transaction). The contract will verify this certificate and transfer the money to the Data Seller in scenarios (a) and (b), or to the Data Buyer in scenario (c).



\end{enumerate}

\subsection{Examples of real-world market participants}
\label{sec:examples-participants}

 
\textbf{Bank credit card transactions.}
Suppose that the Seller is a client of a Bank, who offers on the market his/her (anonymized) credit card transactions. The Buyer can be any entity requiring transactional data to train its machine learning models. 

In this case, the Bank is the ideal Notary since
the Bank can verify that the Seller is actually a client of the Bank, by requiring the Seller to provide information that authenticates him/her.
The Bank can act as a Notary in case of conflict, since it can verify whether the information of credit card transactions sent by the Seller to the Buyer is valid and trustworthy (in particular, by comparing with the Bank's own records of the client's credit card transactions).

\noindent
\textbf{Location data.}
Suppose that the Seller is a client of a Telecommunications company, who offers on the market his/her (anonymized) records with location information. The Buyer can be any entity requiring location data to train its machine learning models. 

In this example, the Telecommunications company (Telco) is the ideal Notary since
the Telco can verify that the Seller is actually a client of the Telco, by requiring the Seller to provide information that authenticates him/her.
The Telco can act as a Notary in case of conflict, and verify whether the location information sent by the Seller to the Buyer is valid and trustworthy (in particular, by comparing with the Telco's own records of the client's location when he/she used mobile phone services).


\section{Conclusion}
\label{sec:conclusion}

Our aim with the Wibson decentralized marketplace presented here is to restore the individuals' ownership over their personal information.
The Wibson protocol will benefit consumers by providing them the ability to control and monetize their personal information. It will also give access to high quality and verified data to organizations which need to train Machine Learning algorithms and models, as well as an explicit consumer consent mechanism which will be absolutely critical as new privacy regulations are coming into effect.

In addition to the marketplace protocol, Wibson also provides primitives (not described here) to solve efficiently the problem of 
fair exchange~\cite{cleve1986limits} by providing an efficient zero-knowledge 
contingent payment mechanism~\cite{campanelli2017zero,sasson2014zerocash,ben2018scalable,bitansky2013succinct}, reminiscent of the 
\emph{secure triggers} cryptographic primitive~\cite{futoransky2006foundations}.

By supporting the principles of transparency, anonymity, fairness and control, we believe that the Wibson marketplace will gain the people's trust 
needed to develop a vibrant data marketplace, that represents a fundamental change in the way organizations collect and use personal information for their data science and business analytics needs.


\end{document}